\documentclass[aps,pra,twocolumn,groupedaddress,showpacs,showkeys]{revtex4-1}

\usepackage{amsmath}
\usepackage{amssymb}
\usepackage{graphicx}
\usepackage{epstopdf}

\usepackage{subcaption}
\usepackage{mwe}

\usepackage{media9}
\usepackage{hyperref}

\begin{document}

\title{Pair interactions of heavy vortices in quantum fluids}
\author{I.A. Pshenichnyuk
\email[correspondence address: ]{ivan.pshenichnyuk@gmail.com}}
\affiliation{Skolkovo Institute of Science and Technology Novaya St., 100, Skolkovo 143025, Russian Federation}
\date{\today}

\begin{abstract}
The dynamics of quantum vortex pairs carrying heavy doping matter trapped inside their cores is studied. The nonlinear classical matter field formalism is used to build a universal mathematical model of a heavy vortex applicable to different types of quantum mixtures. It is shown how the usual vortex dynamics typical for undoped pairs qualitatively changes when heavy dopants are used: heavy vortices with opposite topological charges (chiralities) attract each other, while vortices with the same charge are repelled. The force responsible for such behavior appears as a result of superposition of vortices velocity fields in the presence of doping substance and can be considered as a special realization of the Magnus effect. The force is evaluated quantitatively and its inverse proportionality to the distance is demonstrated. The mechanism described in this paper gives an example of how a light nonlinear classical field may realize repulsive and attractive interactions between embedded heavy impurities.
\end{abstract}      

\maketitle

\section{Introduction}
\label{sec_intro}

Quantum vortices appear in many different physical systems, like trapped dilute Bose-Einstein condensates (BEC) \cite{fetter-2001}, liquid helium \cite{bewley-2006}, exciton-polariton condensates \cite{lagoudakis-2008} and other types of quantum fluids \cite{svistunov}.
Motion and interaction of quantized vortex lines is extensively studied for few decades and many remarkable phenomena including vortex lattices\cite{aboshaeer-2001}, vortex lines reconnection \cite{bewley-2008} and quantum turbulence \cite{barenghi-2014b} were found. Fascinating discoveries were made by investigating quantum mixtures containing two or more interacting components. A number of exotic composite states of matter like skyrmions, coreless vortices \cite{kasamatsu-2005a, saarikoski-2010} and half-quantum vortices \cite{lagoudakis-2009} was discovered to raise the fundamental question of interaction of quantum vortices with matter.

The same question emerges in liquid helium experiments, where various doping substances are used to make vortex cores, which are too thin and optically transparent otherwise, experimentally detectable. For example, in recent experiments vortex lattices in helium nanodroplets \cite{gomez-2014, jones-2016} were doped with xenon atoms which allowed to study them using a femtosecond x-ray coherent diffractive imaging technique. In the experiments of Gordon and colleagues quantum vortices were used to produce long wires of atoms by the ablation of metals in superfluid helium with laser pulses \cite{gordon-2012, gordon-2015}. Many questions were raised in this works that require a detailed understanding of the matter-vortex interaction process, including, for example, the question of unusual distribution of doped vortices inside a droplet.

It was shown theoretically that the scattering of particles on vortices is inelastic \cite{pshenichnyuk-2016}. A certain portion of the particle energy is used to excite Kelvin waves along the vortex core \cite{berloff-2000}. During the interaction the incident particle may be scattered or remain trapped by the vortex depending on the initial amount of energy it possess.
There are certain analogies between this process and the inelastic scattering of electrons on molecules, where an electron may be trapped passing a certain portion of energy to vibrational degrees of freedom of a molecule \cite{pshenichnyuksa-2014, pshenichnyuksa-2006}.
According to helium experiments \cite{gomez-2014, gordon-2015} trapped atoms stick together to produce long cylindrical filaments aligned along the vortex core. The resulting structure is similar to the coreless vortex observed in BEC experiments. A coupled vortex-matter complex, or a doped vortex as it is referred in this paper, is dynamically stable and may demonstrate a distinct behavior, being thus an attractive object for research.

To cover underlying characteristics of vortex-matter interaction which appear in physically different quantum systems a universal theoretical approach is required. Classical complex-valued matter field formalism \cite{svistunov} based on the generalized non linear Schrodinger equation (gNLSE) \cite{berloff-2014} provides the most complete mathematical model of the quantum vortex behavior. It allows to take into account an arbitrary equation of state of a superfluid by incorporating it into the Hamiltonian \cite{berloff-2009, pshenichnyuk-2015} (in the spirit of the density functional theory \cite{ancilotto-2014, ancilotto-2015}). Nonlocal interactions between the particles can be modeled as well \cite{berloff-2000a}. The simplest local form of the equation containing cubic nonlinearity (known as the Gross-Pitaevskii equation) represents a minimalistic model of superfluidity \cite{pitaevskii}. Being a good model for BEC and condensed exciton-polaritons this equation cannot be directly applied to model strongly interacting liquid helium. It is known that only about 10\% of helium is able to condense because of the strong interaction between particles. On the other hand it was demonstrated that NLSE formalism has a certain mathematical parallelism with the Landau two-fluid model and could describe both the condensed phase and the thermal cloud, being thus applicable to model finite temperatures\cite{berloff-2002, berloff-2007}. In this sense liquid helium can be modeled phenomenologically assuming that the initial state is prepared appropriately \cite{berloff-2014}. In its simplest form the classical field equation was used many times in the past to understand qualitatively the behavior of electron bubbles in liquid helium \cite{grant-1974, berloff-2000}. The approach is capable of describing the universal mechanical aspects of matter-vortex interaction in an idealized quantum fluid.

It this paper the dynamics of two doped vortices with different chirality combinations is studied. A special attention is payed to the mass of doping particles which is assumed to be larger than the mass of particles forming a vortex. A doped vortex is called a heavy vortex in this case. It is shown how the doping matter, being heavy, completely changes the character of a vortex pair dynamics.

\begin{figure}
\centerline{\includegraphics[width=0.49\textwidth]{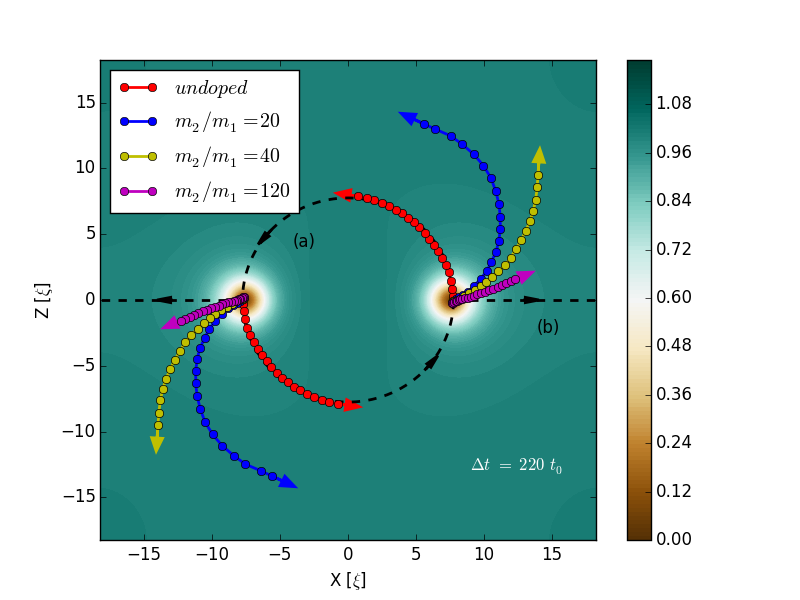}}
\caption {
Motion of two vortices with the same chirality. Density of the host fluid $|\psi(x,0,z)|$ (color coded) is shown for the initial moment of time in the units of $\psi_\infty$. Trajectories of vortices are shown by color arrows which correspond to different dopant masses. Transformation of the circular motion picture (a) typical for pure vortices into the repulsive motion (b) emerging for large relative dopant masses is shown.
\label{fig_pair}}
\end{figure}

\begin{figure}
\centerline{\includegraphics[width=0.49\textwidth]{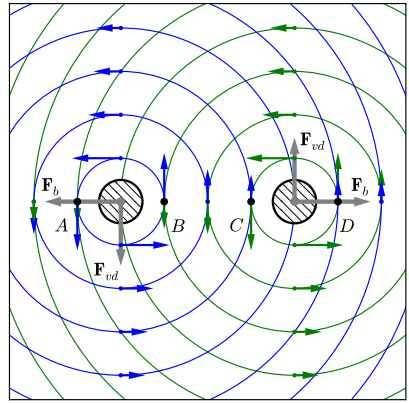}}
\caption {
Schematic explanation of the Bernoulli force acting between two heavy vortices. Velocity fields of vortices are shown using blue and green lines. Velocity decreases with distance $r$ as $1/r$ which is represented by the length of blue and green arrows. Superposition of two fields leads to the situation where the superfluid flow decreases in points B and C and becomes larger in A and D which creates a repulsive force $\mathbf{F}_b$ acting on dopants according to the Bernoulli principle.
\label{fig_bernoulli}}
\end{figure}

\section{Description of the model}
\label{sec_theory}

To study the behavior of doped vortices the following coupled system of nonlinear equations is used 
\begin{equation}
\begin{array}{c}
-i\hbar\psi_t = \frac{\hbar^2}{2m_1}\nabla^2\psi - g_{11}|\psi|^2\psi - g_{12}|\varphi|^2\psi + \mu_1\psi , \\
-i\hbar\varphi_t = \frac{\hbar^2}{2m_2}\nabla^2\varphi - g_{22}|\varphi|^2\varphi - g_{12}|\varphi|^2\varphi + \mu_2\varphi ,
\end{array}
\end{equation}
where the first equation describes the host fluid and the second one defines the behavior of the doping substance. Corresponding complex matter fields $\psi$ and $\varphi$ are assumed to be normalized as $\int|\psi|^2dV=N_1$ and $\int|\varphi|^2dV=N_2$, where $N_1$ and $N_2$ are particle numbers in the fluid and the doping substance respectively and $V$ stands for the volume of the system. Both $\psi$ and $\varphi$ represent cold bosonic matter. Chemical potentials are denoted as $\mu_1$ and $\mu_2$. The equation coefficients are expressed through masses $m_i$ and scattering lengths $l_i$ of the components of the mixture ($i=1,2$) as follows: $g_{11}=4\pi{l_1}\hbar^2/m_1$, $g_{22}=4\pi{l_2}\hbar^2/m_2$, $g_{12}=2\pi{l_{12}}\hbar^2/m_{12}$, where $m_{12}$ is the reduced mass and $l_{12}=(l_1+l_2)/2$.

Practically it is convenient to solve the equations using the units where fields are measured in $\psi_{\infty}=\sqrt{N_1/V}$, distances in  $\xi = \hbar/\sqrt{2m_1g_{11}\psi^2_{\infty}}$ (which is called the healing length) and  time in $t_0=\xi^2m_1/\hbar$ \cite{berloff-2005, berloff-2006}. This units are used throughout the article. It can be shown that the transformed equations contain only relative parameters $m_2/m_1$, $l_2/l_1$ and $N_2/N_1$ and do not depend on the specific type of the host fluid, which makes the results easily scalable to different types of systems. The relative amount of doping matter $N_2/N_1$ is assumed to be small. Moreover, doping particles are assumed to be larger and heavier than particles of the host fluid, so that $N_2/N_1=0.02$, $l_2/l_1=2$. Different values of the relative dopant mass $m_2/m_1$ are considered in the paper. The equations are solved numerically in 3D using a computational box with a volume $V=(74.0\xi \times 36.5\xi \times 74.0\xi)$ and an elementary cell $\Delta{V}=(0.5\xi)^3$ which is enough to study the behavior of a heavy vortex pair.

In the initial state two vortices are created in the host fluid and the doping matter is distributed along their cores in the form of filaments with a cylindrical symmetry (repeating the symmetry of the vortex) and Gauss-like radial distribution of the density. The initial state is then optimized using the imaginary time propagation technique until the steady state is reached. The convergence is controlled by monitoring the time dependence of the full energy of the system. More details about the initial state preparation can be found in the literature \cite{pshenichnyuk-2016, berloff-2000}. During the real time dynamics the vortex-dopant complex moves as a single object if velocities are not too large. Otherwise the heavy filament may decouple, keeping its original form (in the regimes considered in this paper the doping substance and the host fluid remain phase separated also in the absence of vortices). Such behavior is consistent with the available experimental picture. The limits of coupled dynamics are determined by the vortex-dopant binding energy which depends strongly on the volume of the dopant \cite{pshenichnyuk-2016}.

\section{Results and discussion}
\label{sec_results}

\begin{figure}
\centerline{\includegraphics[width=0.49\textwidth]{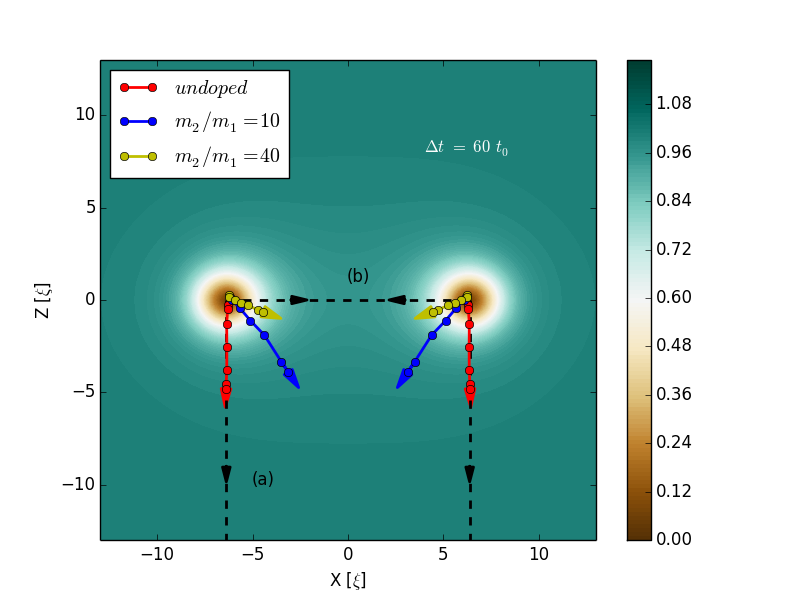}}
\caption {
Motion of two vortices with opposite chiralities. Density of the host fluid $|\psi(x,0,z)|$ (color coded) is shown for the initial moment of time in the units of $\psi_\infty$. Trajectories of vortices during the simulation time ($60t_0$) are shown by color arrows which correspond to different relative dopant masses. Undoped vortices move down along parallel lines (a), as expected.  Attractive force dictated dynamics of type (b) becomes dominant for large relative dopant masses.
\label{fig_antipair}}
\end{figure}

\begin{figure}
\centerline{\includegraphics[width=0.49\textwidth]{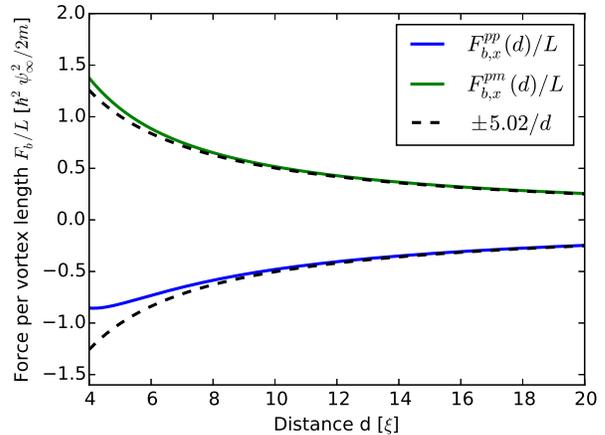}}
\caption {
Bernoulli force responsible for the repulsion and attraction of heavy vortices plotted as a function of inter vortex separation $d$. The green line (attractive case) corresponds to the vortex pair with opposite topological charges, while the blue line (repulsive case) defines the interaction of vortices with the same chirality. A hyperbolic function fit is shown using a black dashed line. The simplified theoretical evaluation based on Eq.~\ref{pair_psi} does not work well at small distances (mind the asymmetry between green and blue curves). At larger distances the force is clearly proportional to ${\pm}1/d$.
\label{fig_forceplot}}
\end{figure}

It is known that undoped vortex pairs behave differently, depending on the combination of vortex winding numbers. Two vortices with the same chirality rotate along a circular trajectory. If chiralities are different vortices move parallel to each other. In both cases the distance $d$ between vortices stays unchanged. Quantum vortex velocity field decays with distance $r$ from its center proportionally to $1/r$ and the linear velocity of the pair motion (expressed in units $\xi/t_0$) is $v=1/d$ \cite{madarassy-2008}.

At the beginning of the numerical experiment two heavy vortices with the same chirality are placed at a distance $d=14\xi$ from each other (both oriented along $y$-axis). Then the system is allowed to evolve during the period of time $\Delta{t}=220t_0$. The initial distribution of the fluid density $|\psi(x,0,z)|$ (color coded) as well as the trajectories of vortices are shown in Fig.~\ref{fig_pair}. Three different masses are considered as well as the undoped vortex case. As it is expected pure vortices move along the circular trajectory (red line), while heavy vortex trajectories strongly depend on the relative dopant particles mass. The system qualitatively changes the character of its motion switching from (a) type trajectories to (b) type (black dashed lines) when $m_2/m_1$ becomes large. 

The force causing the repulsion of vortices can be explained by a special realization of the Magnus effect, responsible for the forces acting on rotating bodies in classical hydro and aerodynamics. The difference is that in the present case the doping matter does not rotate but it is attached to the vortex which produces the rotational velocity field. Since the vortex and the dopant move together one deals with a composite object which possess both a vorticity field and enhanced inertia. Taking into account computational aspects (see below) the term Bernoulli force is used in the paper to name the force.
A visual explanation of the repulsion mechanism is presented in Fig.~\ref{fig_bernoulli}. Velocity fields of two vortices are plotted with blue and green arrows. The length of arrows represents the magnitude of the flow velocity. As a result of superposition of this fields flow velocities in points A and D increase, while velocities in B and C on the contrary decrease. According to the Bernoulli principle, which represents the energy conservation law in hydrodynamics, the pressure (potential energy) is lower in the regions of space where the flow velocity (kinetic energy) is higher. The pressure difference at both sides of each vortex creates a repulsive force $\mathbf{F}_b$ which moves heavy vortices apart.

Another force which defines the dynamics of the pair is a vortex drag force $\mathbf{F}_{vd}$ (see Fig.~\ref{fig_bernoulli}). Super flow, obviously, provides no viscous drag on the impurities, but it moves quantum vortices. Since the vortex is coupled to the dopant with a certain binding energy \cite{pshenichnyuk-2016} it tries to move the dopant along the trajectory (a) in Fig.~\ref{fig_pair}. The actual trajectory is a result of interplay of the Bernoulli  force $\mathbf{F}_b$ and the vortex drag force $\mathbf{F}_{vd}$. The proportion between two forces changes when the mass of the core grows. For heavy vortices the repulsive interaction is mainly dictated by $\mathbf{F}_b$.

The results of another dynamic simulation are shown in Fig.~\ref{fig_antipair}. Starting from the initial state similar to the previous one two heavy vortices with different chiralities are considered. The duration of the simulation is $\Delta{t}=60t_0$ and two different mass regimes are shown. Undoped vortices, as they should, move linearly along black dashed lines (a). When the dopant is added the trajectory changes, deforming continuously from (a) to (b) type while the relative mass $m_2/m_1$ grows. It is easy to show that in the case of different chiralities the mechanism described in Fig.~\ref{fig_bernoulli} generates the attractive force $\mathbf{F}_b$. The conclusion that follows is that in the case of opposite topological charges (winding numbers) heavy vortices attract each other, while in the case of the same charges the repulsive interaction appears. If dopants are heavy, the repulsive-attractive behavior dominates over the rotational motion and parallel motion typical for undoped vortices.

As vortices move closer to each other their velocity becomes larger (similar to the vortex rings which move faster for smaller radius). At a certain distance, when the critical velocity is reached, the vortex pair decouples and moves separately from the doping filaments. According to the obtained results decoupling happens approximately at $d=10$ for $m_2/m_1=40$ and at $d=5$ for $m_2/m_1=10$. Since $v=1/d$, corresponding decoupling velocities are $v_{dec}=0.1$ and $v_{dec}=0.2$. Or, in other words, for heavier dopants decoupling occurs at lower velocities. 

An approximate evaluation of the force $\mathbf{F}_b$ which governs the dynamics of heavy vortices can be obtained. Matter field $\psi$ corresponding to the single vortex placed at $x=z=0$ and oriented along $y$-axis is well fitted by the following approximate relation \cite{parks-1966}
\begin{equation}
  \psi(x,y,z) = \frac{\psi_{\infty}(x+iz)}{\sqrt{x^2+z^2+\xi^2}} .
  \label{fetter}
\end{equation}
The change of chirality corresponds to the complex conjugation of Eq.~\ref{fetter}. For a pair of vortices separated by distance $d$ one gets
\begin{equation}
  \psi_{p} = \psi_1(x,y,z)\psi_2(x-d,y,z)/\psi_{\infty} ,
  \label{pair_psi}
\end{equation}
where each factor is given by Eq.\ref{fetter}. Assuming that a cylindrical doping filament with radius $R$ and length $L$ (coinciding with the length of the vortex) is placed inside the vortex core of the first vortex at $x=z=0$ one may compute the force acting on the cylinder which is caused by the velocity field given by Eq.\ref{pair_psi}. The field perturbation caused by the dopant is neglected. The expression for the Bernoulli force \cite{sergeev-2006, pshenichnyuk-2016} reads
\begin{equation}
  \mathbf{F}_b = \int\limits_S K(\mathbf{r}) \mathbf{n}(\mathbf{r}) dS
  \label{bernoulli}
\end{equation}
where $K(\mathbf{r})=\rho{v^2}/2$ is a fluid kinetic energy density, $\mathbf{n}(\mathbf{r})$ is a unit vector normal to the surface of the cylinder $S$ and the integral is taken over this surface.

Assuming the complex matter field in the form $\psi\equiv|\psi|e^{i\phi}$ for a superfluid density and velocity one gets: $\rho=|\psi|^2m_1$, $\mathbf{v}= \frac{\hbar}{m_1}\nabla\phi$. Using Eq.\ref{fetter} and Eq.\ref{pair_psi} the following expressions for the kinetic energy density on the surface of the cylinder can be derived
\begin{equation}
K^{pp} = 
\frac{\hbar^2\psi^2_{\infty}(4R^2 + d^2 - 4Rd\cos{\alpha})}
{2m_1(R^2+\xi^2)(R^2+d^2+\xi^2-2Rd\cos{\alpha})} ,
\end{equation}
\begin{equation}
K^{pm} = 
\frac{\hbar^2\psi^2_{\infty}d^2}{2m_1(R^2+\xi^2)(R^2+d^2+\xi^2-2Rd\cos{\alpha})} ,
\end{equation}
where $\alpha$ is an azimuthal angle in cylindrical coordinates associated with the dopant. Upper indices are used to denote "plus-plus" and "plus-minus" topological charges combinations for a pair. The integral in Eq.~\ref{bernoulli} now can be easily evaluated numerically in cylindric coordinates where $dS = LRd\alpha$ and $\mathbf{n}(\mathbf{r}) = (\cos\alpha,\sin\alpha)$.

The force $\mathbf{F}_b$ responsible for the repulsion and attraction of heavy vortices is visualized in Fig.\ref{fig_forceplot}. The radius of the doping filament $R=2\xi$ is taken. The force acting on the vortex of a unit length is plotted as a function of inter vortex distance $d$ (distances are measured in healing lengths $\xi$). As it is expected the force acts along the line connecting vortices (along $x$ in the used coordinates), while other components are equal to zero. The force is repulsive (blue line) in a case when two vortices with the same chirality interact and attractive (green line) when chiralities are different. The obtained curves are well fitted with the hyperbolic function $f(d)=\pm5.02/d$ which is plotted using the black dashed line. Approximations which were introduced to evaluate the force obviously can not be applied at small distances. Practically, asymmetry between repulsive and attractive forces becomes noticeable at $d<10$. On the other hand, for heavy regimes ($m_2/m_1=40$) vortices decouple from dopants at distances of the order of 10, which means that the derived force does not exist when $d<10$.

At larger distances the force decays proportionally to $1/d$ (similar to the velocity field of the vortex). Mechanical computations based on the defined force (and neglecting the vortex drag force) give good evaluation for the radial movement of vortices in heavy regimes. The obtained result may be applicable to describe a certain class of phenomena involving heavy vortices. The existence of repulsive force, for instance, should influence the structure of vortex lattices, containing heavily doped vortices. It can be the case in recent helium nanodroplets experiments \cite{jones-2016}.

\section{Conclusion}
\label{sec_conclusion}

The dynamics of quantum vortices doped with foreign particles is studied. It is demonstrated that the character of the vortex pair motion changes qualitatively when the mass of doping particles becomes large compared to the mass of fluid particles. In heavy regimes the dynamics transforms from rotation and parallel motion to repulsion and attraction, depending on the combination of chiralities of vortices. The observed behavior is caused by the competition of the vortex drag force and Bernoulli force. The latter one becomes dominant for heavy vortices. The mechanism producing the Bernoulli force is explained through the  inhomogeneity of the pressure field caused by the superposition of heavy vortices velocity fields. It is shown that the force is inversely proportional to the distance $d$ between heavy vortices. The obtained results give an example of how a nonlinear complex valued classical field may act as a repulsive and attractive (depending on the combination of topological charges) force generator between heavy particles. The Bernoulli force should be taken into account during the interpretation of experimental data on quantum turbulence. It may influence the structure of vortex lattices in doped helium nanodroplets.

\section*{Acknowledgment}

Stimulating discussions with Natalia Berloff and Andrey Vilesov are greatly acknowledged.

\bibliography{paper_heavypairs}

\end{document}